\newcommand{\bfx}{\ensuremath{{\bf x}}}

\newcommand{\bfr}{\ensuremath{{\bf r}}}
\newcommand{\bfs}{\ensuremath{{\bf s}}}
\newcommand{\bfk}{\ensuremath{{\bf k}}}

\newcommand{\bfn}{\ensuremath{{\bf n}}}
\newcommand{\conv}{\ensuremath{\star}}
\newcommand{\Vmax}{V_{\rm max}}

\newcommand{\bfM}{\ensuremath{{\bf M}}}
\newcommand{\bfI}{\ensuremath{{\bf I}}}
\newcommand{\bfA}{\ensuremath{{\bf A}}}

\documentclass[useAMS,usenatbib]{mn2e}

\usepackage{placeins}

\usepackage{graphicx} 
\usepackage{multirow}
\usepackage{array}

\usepackage{amsmath,amssymb}

\usepackage[T1]{fontenc}
\usepackage[latin9]{inputenc}
\usepackage{float}
\usepackage{graphicx}
\usepackage{esint}

\usepackage{hyperref}

\makeatletter
\DeclareFontEncoding{LGR}{}{}
\DeclareTextSymbol{\~}{LGR}{126}
\begin{document}

\title[Accelerating 2 and 3PCF with FTs]{Accelerating the 2-point and 3-point galaxy correlation functions using Fourier transforms}

\author{\makeatauthor}

\author[Slepian and Eisenstein]{Zachary Slepian\thanks{E-mail: zslepian@cfa.harvard.edu} and Daniel J. Eisenstein\thanks{E-mail: deisenstein@cfa.harvard.edu}\\
Harvard-Smithsonian Center for Astrophysics, Cambridge, MA 02138\\
}
\maketitle

\begin{abstract}
Though Fourier Transforms (FTs) are a common technique for finding correlation functions, they are not typically used in computations of the anisotropy of
the two-point correlation function (2PCF) about the line of
sight in wide-angle surveys because the line-of-sight direction
is not constant on the Cartesian grid. Here we show how FTs can be used to compute the multipole moments of the anisotropic 2PCF. We also show how FTs can be used to accelerate the 3PCF algorithm of Slepian \& Eisenstein (2015). In both cases, these FT methods
allow one to avoid the computational cost of pair counting, which
scales as the square of the number density of objects in the
survey. With the upcoming large datasets of DESI, Euclid, and LSST, FT techniques will therefore offer an important complement to simple pair or triplet counts.
\end{abstract}

\begin{keywords}
 cosmology: large-scale structure of Universe, methods: data analysis, statistical
\end{keywords}

\section{Introduction}
\label{sec:intro}
In studying the large-scale clustering of galaxies, we commonly
use the two and three-point correlation functions $\xi$ (2PCF) and $\zeta$ (3PCF) (Peebles 1980; Bernardeau et al. 2002; Szapudi 2005 for reviews; recent observational work on the 2PCF is Anderson et al. 2014; on the 3PCF, Kayo et al. 2004; McBride et al. 2011a, b; Guo et al. 2014). These are correlations of the fractional overdensity field $\delta(\bfx)
= \rho(\bfx)/\bar\rho - 1$, where $\rho(\bfx)$ is the density field
and $\bar\rho$ is the mean density. Also commonly used is the anisotropic 2PCF $\xi_{\rm aniso}$, often written as multipole moments of the 2PCF with respect to the line of sight (Cabr\'e \& Gazta\~{n}aga 2009; Okumura \& Jing 2011; Reid et al. 2012; Samushia, Percival \& Raccanelli 2012; Chuang \& Wang 2012; Chuang \& Wang 2013a, b; Chuang et al. 2013; S\'anchez et al. 2013; Xu et al. 2013; Ross, Percival \& Manera 2015; White et al. 2015 for recent modeling and observational work). Typically, direct counting is
used to compute $\xi$, $\xi_{\rm aniso}$, and $\zeta$. The calculation of $\xi$ and $\xi_{\rm aniso}$
scales as $Nn \Vmax$, while that of $\zeta$ scales as $N(n\Vmax)^2$, where
$N$ is the number of galaxies in the survey, $n$ is the number
density, and $\Vmax$ is the volume of a sphere of the maximum radius out to which the correlation is measured.  For
surveys such as the Sloan Digital Sky Survey (SDSS) (Alam et al. 2015 for latest release), with of order
one million galaxies, and upcoming efforts like Euclid (Laureijs et al. 2011), Large
Synoptic Survey Telescope (LSST) (LSST Dark Energy Science Collaboration 2012), Dark Energy Spectroscopic Instrument
(DESI) (Levi et al. 2013), and WFIRST (Spergel et al. 2013) (see also Jain et al. 2015), with tens of millions to billions of galaxies, these scalings
render the 2PCF computationally expensive and the 3PCF computationally
prohibitive for large $\Vmax$.
Therefore, techniques that scale more favorably with number and number
density of objects merit consideration. In this
work, we argue for the utility of Fourier methods in measuring the two and three-point correlation functions,
as their computational cost depends primarily on the number of grid
cells $N_g$ into which the survey volume is discretized (though gridding the density field depends linearly on the number of
galaxies).  While the methods will introduce
artifacts due to the grid resolution, the balance of performance versus
accuracy may be desirable for some applications, for example the 
analysis of large-scale correlations in large numbers of mock catalogs.\\
\indent It has long been known that the 2PCF can be computed using
FTs, but, for reasons we detail further below (Section \ref{sec:pair_vs_ft}), this
has not been the favored approach.  It has been less appreciated
that the anisotropic 2PCF and 3PCF can also be computed
using FTs, though they have been used for the projected 3PCF (Zheng 2004) and the 3PCF of the Cosmic Microwave Background (Chen \& Szapudi 2005), as well as multipole moments of the power spectrum (Bianchi et al. 2015; Scoccimarro 2015) and bispectrum (Scoccimarro 2015). Though this will not be our focus here, FTs are standardly used to compute the power spectrum, typically with FKP weighting (Feldman, Kaiser \& Peacock 1994). A recent implementation is discussed in Anderson et al. (2014) and Percival et al. (2014).\\
\indent In this paper, we begin by showing that several perceived disadvantages of using FTs
for the standard 2PCF can be avoided with a particular way of setting
up the Landy-Szalay estimator (Section \ref{sec:pair_vs_ft}). 
We then show that the anisotropic 2PCF can also be
straightforwardly handled with FTs (Section \ref{sec:aniso_corr}).
In Section \ref{sec:3pcf_fts}, we show that the 3PCF algorithm recently presented in Slepian
\& Eisenstein (2015; hereafter SE15), which already offers roughly
a factor of 500 speed improvement relative to naive triplet counting,
can in many circumstances be further accelerated using FTs. Section \ref{sec:DandC} concludes.

\section{Pair counting with Fourier Transforms}
\label{sec:pair_vs_ft}
\subsection{Estimator and passage to the FT}
\label{sec:subsec_est_to_ft}

It is well-known that the 2PCF of
a field can be computed quickly using the FT: it is simply the convolution of the density field with itself, and hence by the Convolution Theorem reduces to multiplication in Fourier space:
\begin{equation}
\xi(\bfr) \equiv \int d\bfx\ \delta(\bfx) \delta(\bfx+\bfr) 
	= \int {d\bfk\over (2\pi)^3} |\tilde\delta(\bfk)|^2 e^{-i\bfk\cdot\bfr}
\end{equation}
where here and throughout $\tilde\delta(\bfk) = \int d\bfx\ \delta(\bfx) e^{i\bfk\cdot\bfx}$
is the FT of the field.  Numerically, the integrals become sums
over wavenumbers quantized to fit in the finite box on which the
integral is evaluated, and for periodic boundary conditions the
largest wavelength is simply the periodicity. For a non-periodic
survey, we can still use this method by expanding the grid to be
well larger than the survey and setting $\delta=0$ outside the
survey volume.

However, this is not usually the method employed in cosmology, particularly in wide-field surveys.  Instead, one typically uses an explicit counting of 
pairs, such as the Landy-Szalay (1993) estimator,
\begin{align}
\hat{\xi}({\bf S}) & =\frac{NN}{RR}=\frac{\int d \bfr_1 d\bfr_2\; \theta(\bfr_1-\bfr_2;{\bf S}) N(\bfr_1)N(\bfr_2)}{\int d\bfr_1 d\bfr_2\;\theta( \bfr_1-\bfr_2; {\bf S}) R(\bfr_1)R(\bfr_2)},
\label{eqn:LS_est_short}
%\label{eqn:LS_NR}
\end{align}
with $N\equiv(D-R)$, $D$ the data, $R$ the random counts, and $\theta$ a binning function that is non-zero where its first argument is within a three-dimensional volume labeled by ${\bf S}$. In this Section we focus on bins of fixed Cartesian separation. The random catalog is chosen to be a Monte
Carlo realization of the mean density field, with weights such that
the mean densities of the randoms and the data match. 
Changing variables to the separation $\bfs \equiv \bfr_2-\bfr_1$, we have
\begin{align}
\hat{\xi}({\bf S}) &=\frac{\int d\bfs \; \theta( \bfs; {\bf S}) \int d\bfr_1  N(\bfr_1)N(\bfr_1+\bfs)}{\int d\bfs \; \theta( \bfs; {\bf S}) \int d\bfr_1  R(\bfr_1)R(\bfr_1+\bfs)}\nonumber\\
& = \frac{\int d\bfs \;\theta(\bfs; {\bf S})[N\conv N](\bfs)}{\int d\bfs \;\theta(\bfs; {\bf S})[R\conv R](\bfs)}.
\end{align}
Thus the binned estimator is simply a binned convolution, and the convolution can be evaluated with FTs in $\mathcal{O}(N_g\log N_g)$ time (Cooley \& Tukey 1965; Press et al. 2007).  Forming the
gridded density field the FTs require is linear in the number of particles
(e.g., using triangular cloud-in-cell interpolation (Hockney \& Eastwood 1981; Jing 2005)),
so we avoid any quadratic scaling with the galaxy density. Therefore, in principle the Landy-Szalay estimator can be straightforwardly evaluated using FTs.
 
\subsection{Comparing FTs and pair counting}
\label{sec:subsec_pair_vs_ft}
 We now discuss the practical disadvantages and advantages of pair counting relative to a Fourier approach.  Ultimately we will show that several perceived issues with the FT method that seemed to make pair counting more favorable can be resolved by the approach of this paper.
 
Relative to the Fourier method, pair counting has the disadvantage 
that the work scales as the square of the number of points.
It is a simple optimization, using trees or grids, to avoid working
with pairs that are much further separated than the maximum scale
of interest.  However for work on large scales, the large number of
pairs can be computationally burdensome.  This is particularly true
because the number of randoms typically should be much
larger than (of order 100 times) the number of data points.
The computational expense can be somewhat reduced by using tree methods
that aggregate many points into cells that are then added to the count
of a given separation bin as a unit. However, this only helps if the binning
is substantially coarser than the interparticle spacing, which often is not
the case in practical galaxy surveys.

Pair counting does have some important advantages. First, it avoids any
gridding of the data.  Gridding results in small displacements of the
effective particle positions, which in turn produce correlation
results that are smoothed versions of the true answer (see Jing 2005 for discussion of the effect in the power spectrum and how it can be ameliorated).  Cartesian
gridding can also cause artifacts when summing over separation bins
in spherical coordinates.  Decreasing the grid spacing can decrease
these biases, but at an increased computational cost.
Second, pair counting allows easy computation of the anisotropy of the 
correlations relative to the line of sight, since in real space the orientation of each pair to the line of sight is clear. Fourier methods, on the other hand, use a Cartesian basis
that treats positions in the survey without preference as to orientiation to the line of sight, appearing to destroy
this information.  Third, pair counting produces an unbiased estimate
of the correlation function regardless of the survey geometry. In contrast, an FT-based convolution of the $\delta$ field
yields a misnormalized result due to the zero-padding outside of
the survey volume.

In this work, we show that the last two of these problems can be
easily avoided when using FTs. Here, we discuss the zero-padding issue, deferring 
the anisotropic correlations to Section \ref{sec:aniso_corr}. Equation (\ref{eqn:LS_est_short}) shows that there is no need to form the $\delta$ field.  The zero padding of the grid beyond the
survey boundary here is of no consequence because it will enter
both the numerator and denominator of equation (\ref{eqn:LS_est_short})
and hence cancel out. Thus the value of $\hat{\xi}$ should be the same, up
to grid smoothing, whether one has used pair counting or the FT.
Consequently any advantage of the Landy-Szalay computation of the
monopole of the correlation function, e.g., as regards the integral
constraint (Coil 2012 for definition), will remain.

Finally, we note that there is a common misunderstanding that
for non-periodic volumes one must embed in a periodic domain that
is twice the size of the original survey.  This is not true if one
is only interested in a limited range of separations $\bfs$.  One
need only use a periodic embedding large enough that the separation
between any point in the survey and any point in the periodic
duplicate is larger than $|\bfs|$.

\section{Anisotropic correlations}
\label{sec:aniso_corr}

%We now consider the case in which the bin $\theta$ is a 
%function of position $\bfx$ in the survey as well as separation $\bfs$, i.e. $\theta=\theta(\bfx, \bfs ; \bfr)$.  
%We are particularly concerned with the case in which $\theta$ is
%separable
%\begin{equation}
%\theta(\bfx, \bfs ; \bfr) = \sum_i f_i(\bfx) \theta_i(\bfs ; \bfr).
%\label{eqn:sep_window_fn}
%\end{equation}
We now turn to the anisotropic 2PCF, described in terms of its multipole moments $\xi_\ell$.  The anisotropies of the correlation functions, and most importantly
the quadrupole $\xi_2$, have important cosmological purpose for the
measurement of the Alcock-Paczynski effect (Alcock \& Paczynski 1979), redshift-space distortions (Kaiser 1987; Hamilton 1998 for a review), and anisotropic baryon acoustic oscillation (BAO) signature (Gazta\~{n}aga, Cabr\'e \& Hui 2009). 

Here the multipole moments are with respect to the angle between the pair separation $\bfs$ and the line of sight $\bfn = (\bfr_1+\bfr_2)/2$, and we define $\mu=\hat{\bfs}\cdot \hat{\bfn}$. We first write $\xi_{\rm aniso}$ as a function of bin $S$ in separation magnitude $s=|\bfs|$ and $\mu$:
\begin{align}
\xi_{\rm aniso}(S,\mu)=\frac{\mathcal{N}(S,\mu)}{\mathcal{R}(S,\mu)},
\label{eqn:xi_N_R}
\end{align}
where $\mathcal{N}$ and $\mathcal{R}$ respectively denote $NN$ and $RR$. We expand $\xi_{\rm aniso}$, $\mathcal{N}$, and $\mathcal{R}$ as multipole series:
\begin{align}
\mathcal{\xi_{\rm aniso}}(S,\mu)&=\sum_{\ell=0}^{\infty} \xi_{\ell}(S)P_{\ell}(\mu),\;\;\;\mathcal{N}(S,\mu)=\sum_{k=0}^{\infty} \mathcal{N}_k(S)P_k(\mu),\nonumber\\
&\mathcal{R}(S,\mu)=\sum_{j=0}^{\infty} \mathcal{R}_{j}(r)P_{j}(\mu).
\label{eqn:N_and_R_series}
\end{align}
$P_{\ell}$ is a Legendre polynomial, and as usual, these relations imply $\xi_{\ell}(S) = [(2\ell+1)/2] \int d\mu P_\ell(\mu) \xi_{\rm aniso}(S,\mu)$
and similarly for $N_k(S)$ and $R_j(S)$. Multiplying equation (\ref{eqn:xi_N_R}) through by $\mathcal{R}$ and then inserting equations (\ref{eqn:N_and_R_series}), we find
\begin{align}
\sum_{lj}\xi_{\ell}(S)\mathcal{R}_j (S) P_{\ell}(\mu) P_{j}(\mu)=\sum_k \mathcal{N}_k(S)P_k(\mu).
\end{align}
Using a linearization formula for the product of two Legendre polynomials we obtain, with arguments suppressed, 
\begin{align}
\sum_{\ell j q} \mathcal{\xi}_{\ell} \mathcal{R}_{j}(2j+1)\left(\begin{array}{ccc}
\ell & j & q\\
0 & 0 & 0
\end{array}\right)^{2} P_q=\sum_{k}\mathcal{N}_k P_k.
\end{align}
The Wigner 3j-symbol above describes angular momentum coupling. Using orthogonality, separating out the $j=0$ term, dividing through by $\mathcal{R}_0$, and defining $f_{j}=\mathcal{R}_{j}/\mathcal{R}_0$, we obtain
\begin{align}
\frac{\mathcal{N}_k}{\mathcal{R}_0}=\xi_{k} +\sum_{\ell} \xi_{\ell} (2k+1) \sum_{j>0} \left(\begin{array}{ccc}
\ell & j & k\\
0 & 0 & 0
\end{array}\right)^{2} f_{j}.
\label{eqn:edge_corrxn_1}
\end{align}
Defining a multipole coupling matrix $\bfM$ with elements
\begin{align}
M_{k \ell}=(2k+1)\sum_{j>0}\left(\begin{array}{ccc}
\ell & j & k\\
0 & 0 & 0
\end{array}\right)^{2}f_{j}
\end{align}
we see that equation (\ref{eqn:edge_corrxn_1}) can be written as
\begin{align}
\vec{\mathcal{N}}/\mathcal{R}_0=(\bfI+\bfM)\vec{\xi}_{\rm aniso}\equiv \bfA \vec{\xi}_{\rm aniso},
\label{eqn:edge_corrxn_2}
\end{align}
where $\vec{\mathcal{N}}=(\mathcal{N}_0,\mathcal{N}_1,\cdots,\mathcal{N}_{\ell_{\rm max}})$ and analogously for $\vec{\xi}_{\rm aniso}$.
The edge-correction equation (\ref{eqn:edge_corrxn_2}) can then be
solved by matrix inversion.  Formally we need all multipoles
of the randoms $\mathcal{R}_{j}$ to obtain the solution, but in
practice the $f_{j}$ should fall off so quickly that measuring
only out to some $\ell_{\rm max}$ and using this to invert a truncated,
finite-dimensional sub-matrix of $\bfM$ should suffice. For more
detailed discussion of similar issues arising in the 3PCF, see SE15 Section 4.2. Note also that if we are computing an auto-correlation, then by parity all odd-order coefficients in equation (\ref{eqn:N_and_R_series}) vanish.

%we seek only $\xi_2$, we set $k=2$ in equation (\ref{eqn:edge_corrxn_1}). This, combined with
%the constraint on the 3j-symbols that the top row has an even sum
%and the fact that odd moments of $\xi_{\rm aniso}$ vanish, means that we only
%require $f_{l'}$ for $l'$ even.

With equation (\ref{eqn:edge_corrxn_2}) for the vector of multipole coefficients $\vec{\xi}_{\rm aniso}$ in hand, our task now becomes measuring the $\mathcal{N}_k$ and $\mathcal{R}_{j}$ it requires. We restrict to the case where the pairs project to only a small angle on the sky; for discussion of wide-angle effects, see Samushia, Percival \& Raccanelli (2012) and Raccanelli et al. (2013). In this limit, we then approximate $\mu=\hat{\bf s}\cdot\hat{\bf n}$ by $\hat{\bf s}\cdot\hat{{\bf r}_1}$, i.e., approximating that the
line of sight to the pair is very nearly the line of sight to one
member (Yamamoto et al. 2006).  We
write
\begin{align}
\mathcal{N}_k(S)=&(2k+1)\int s^2ds\;\theta(s;S)\int r_1^2dr_1\nonumber\\
& \times \int d\Omega_s d\Omega_{r_1} P_k(\hat{\bfs}\cdot\hat{\bfr}_1)N(\bfr_1)N(\bfr_1+\bfs)
\end{align}
and analogously for $\mathcal{R}_{j}$. The integral over $dr_1$ averages over translations. Consolidating integrals, we find
\begin{align}
\mathcal{N}_k(s)=(2k+1)\int d\bfs \;\theta(s;r) \int d\bfr_1 \; P_k(\hat{\bfs}\cdot\hat{\bfr}_1)N(\bfr_1)N(\bfr_1+\bfs).
\end{align}
We now show how to cast the inner integral as a convolution, which will then permit its evaluation via FTs. Using the spherical harmonic addition theorem (Arfken, Weber \& Harris 2013, hereafter AWH13, equation 16.57) the integral becomes
\begin{align}
&\mathcal{N}_k(s)=4\pi \sum_{m=-k}^{k}\int d\bfs\;\theta(s;S)\nonumber\\
&\times \int d\bfr_1 \;Y_{k m}(\hat{\bfs})Y_{k m}^*(\hat{\bfr}_1)N(\bfr_1)N(\bfr_1+\bfs)\nonumber\\
&=4\pi \sum_{m=-k}^{k} \int d\bfs \;\theta(s;S)Y_{k m}(\hat{\bfs})[(NY_{k m}^*) \conv N](\bfs).
\end{align}
The approach here generalizes to any separable kernel inserted in place of the $P_{\ell}$ above.

With the problem thus set up as a convolution, to compute a particular
multipole $\mathcal{N}_k$ we will need $2\ell+2$ real forward FTs, one for
the data minus randoms, $N$, and then $2\ell+1$ for the independent components 
of the $NY_{\ell m}$.
We then need $2\ell+2$ real inverse transforms after taking the products in
Fourier space.  
Note that once the convolution is computed, we can perform all of the
integrals over the binning as needed.  Further, we can separate the
real and imaginary components of the spherical harmonics and compute all of these
$2\ell+1$ terms sequentially, which allows us to store only 3 copies
of the full grid at a time ($N$ and its FT, plus the working space for the 
convolution), while accumulating the resulting contributions to $\mathcal{N}_k$. 

The computation of $\mathcal{R}_{j}$ proceeds identically.  However,
we note that because the $\mathcal{R}$ pair count does not involve a near-cancellation
as $\mathcal{N}=(D-R)^2$ does, one can use a substantially smaller random catalog
when computing $\mathcal{R}_{j}$.  Inspecting equation (11) shows
that $\mathcal{R}_0$ appears as a normalization of the correlation
function, while the $f_{j}$ ratios only slightly mix terms.
Moreover, when computing repeatedly on large numbers of mock catalogs,
one would generally not need to repeat the computation of $\mathcal{R}$
for each.
\iffalse
Finally, recall that we needed $\mathcal{N}_k$ to solve equation (\ref{eqn:edge_corrxn_2}); we also need $\mathcal{R}_{l'}$, and we need both out to $l_{\rm max}$.  Thus the total number of FTs required is $2[1+2\sum_{\ell=0}^{\ell_{\rm max}}(2\ell+1)]=2+4(\ell_{\rm max}+1)^2$; 1 for $N$ and $R$ forward each, the sum for all multipoles from zero to $\ell_{\rm max}$, and a factor of 2 accounting for the forward and inverse and another for the $N$ and the $R$.
\fi
%%% The forward and inverse transforms of
%%% $nf_i$ and $n$ need only be done once since they do not depend on
%%% the radial bin (it is implicit in $\theta_i$); once the convolution
%%% in equation (\ref{eqn:conv_xi2}) is obtained, we can perform as
%%% many integrals over the binning (i.e. over $\bfs$) as needed.  
%%% Thus we need a total of $4\ell+3$ FTs. 
%%% Further, the radial binning greatly reduces
%%% the work of summing over the $\theta_i$ windows since they will be
%%% non-zero over only a small portion of the grid.

%Why did DE write one gets the monopole out of this? I don't see it because Legendre basis is orthogonal.

\section{3PCF with Fourier Transforms}
\label{sec:3pcf_fts}
We now show how to use FTs to accelerate the algorithm for measuring multipole moments of the 3PCF presented in SE15.  Note that here we consider only the isotropic 3PCF and do not track orientiation to the line of sight. We first recall that this algorithm measures the binned radial coefficients in an expansion of the 3PCF as
\begin{align}
\zeta(S_1,S_2;\hat{\bfr}_1\cdot \hat{\bfr}_2) = \sum_{\ell}\zeta_{\ell}(S_1,S_2)P_{\ell}(\hat{\bfs}_1\cdot \hat{\bfs}_2).
\label{eqn:zeta_def}
\end{align}
The algorithm exploits the spherical harmonic addition theorem to break the Legendre polynomial into a separated sum of spherical harmonics and then obtains the spherical harmonic coefficients $a_{\ell m}(S;\bfx)$ in each radial bin (designated by $S$) and around each galaxy in the survey (with position $\bfx$).  These are combined locally about each galaxy to form the binned estimate about that origin,
\begin{align}
&\hat{\zeta}_{\ell}(S_1,S_2;\bfx)=\frac{1}{4\pi}\delta(\bfx)\sum_{m=-\ell}^l a_{\ell m}(S_1;\bfx)a_{\ell m}^*(S_2;\bfx)
\label{eqn:zetal_ito_alms}
\end{align}
(SE15 equation 15), and finally averaged over all possible origins $\bfx$ to find
\begin{align}
\zeta_{\ell}(S_1,S_2)&=\frac{1}{V}\int d\bfx \;\hat{\zeta}_{\ell}(S_1,S_2;\bfx)
\end{align}
(SE15 equation 12).

Here we show that computing the local $a_{\ell m}(S;\bfx)$ is simply a convolution and so can be accelerated with FTs. We begin with SE15 equation 14 for the $a_{\ell m}$ about a particular origin of an arbitrary density field $\delta$:
\begin{align}
a_{\ell m}(S;\bfx) =\int d\Omega' \; Y_{\ell m}^*(\hat{\bfr}')\int r'^2 dr'  \; \theta(|\bfr'|;S)\delta(\bfr'+\bfx).
\label{eqn:SE15_14}
\end{align}
Consolidating integrals, this becomes
\begin{align}
a_{\ell m}(S;\bfx)  =\int d\bfr' \; Y_{\ell m}^*(\hat{\bfr}')\theta(|\bfr'|;S)\delta(\bfr'+\bfx),
\label{eqn:SE15_14_consol}
\end{align}
which clearly has the form of a convolution. Hence by the Convolution Theorem
\begin{align}
a_{\ell m}(S;\bfx)={\rm FT}^{-1}\left\{ \tilde{K}_{\ell m}(\bfk ; S)\tilde{\delta}(\bfk)  \right\}(\bfx),
\end{align}
where again $\tilde{\delta}(\bfk)$ is the FT of the density field $\delta(\bfr')$ and $\tilde{K}_{\ell m}(\bfk;S)$ is the FT of the kernel
\begin{align}
K_{\ell m}(\bfr';S)\equiv Y_{\ell m}^*(\hat{\bfr}')\theta(|\bfr'|;S).
\label{eqn:K_lm}
\end{align}
Thus where in SE15 we needed an $\mathcal{O}(n \Vmax)$
operation about each possible origin $\bfx$ to compute each $a_{\ell m}$ for
a given radial bin, and hence $\mathcal{O}(Nn \Vmax)$ operations total
for each radial bin, we now simply need $N_{g}\log N_g$ operations
total to compute each $a_{\ell m}$ in a given radial bin and for ${\it
all}$ origins. 

In detail, in SE15 the $\zeta_{\ell}$ are expressed in terms of multipole moments of the $NNN$ and $RRR$ fields (see SE15 Section 4), which can be obtained using equations (\ref{eqn:SE15_14})-(\ref{eqn:K_lm}) with $\delta =NNN$ and then $\delta =RRR$ successively.  If we want the multipole coefficients of, e.g., $NNN$, up to $\ell_{\rm max}$ in $N_{\rm bins}$ radial bins, we need one real forward FT of
the density field, $N_{\rm bins}(\ell_{\rm max}+1)^2$ real forward transforms for the kernels $K_{\ell m}$, and
finally this same number of real inverse transforms after taking products
in Fourier space. The same holds for $RRR$.
%\todo{DJE needs to check this.}

The kernel $K_{\ell m}$ is simple and so its forward FTs can be done analytically,
essentially halving the total number of FTs required. We have
\begin{align}
\tilde{K}_{\ell m}(\bfk;S)&=\int d\bfr' e^{i\bfk\cdot\bfr'} \theta(|\bfr'|;S) Y_{\ell m}^*(\hat{\bfr}')\nonumber\\
&=(4\pi) i^{\ell} Y_{\ell m}^*(\hat{k}) \bar{j}_{\ell}(k;S).
\label{eqn:analytic_klm}
\end{align}
We expanded the plane wave using AWH13 equation 16.61, performed the angular integral by orthogonality, and defined
\begin{equation}
\bar{j}_{\ell}(k;S)=\int u^2du  j_{\ell}(ku)\theta(u;S).
\label{eqn:jlbar}
\end{equation}
Analytically evaluating the kernel's FTs is not always favorable. Doing the
FTs all numerically treats data and kernel on the same footing:
both will be gridded using, e.g., cloud-in-cell and then transformed.
If one computes $\tilde{K}_{\ell m}$ analytically, however, one
must then grid in Fourier space so as to match the gridded, transformed
data.  Transforming and then gridding only reduces to gridding and
then transforming in the limit of a very fine grid. Otherwise we
expect this reordering of operations might introduce additional
artifacts.   However it is
precisely in applications where the grid is extremely fine that the
FT will take longest and so imply the greatest need to reduce the
number of transforms by analytic evaluation of $\tilde{K}_{\ell m}$.  

\section{Discussion and Conclusions}
\label{sec:DandC}

We have presented Fourier Transform methods for the computation of the
2PCF, anisotropic 2PCF, and 3PCF.  For the 2PCF, we have shown that the familiar Landy-Szalay
estimator can be immediately translated into an FT computation.  We
then show that the multipoles of the anisotropic 2PCF can be computed
by FTs, despite the curvature of the sky relative to the Cartesian
grid.  After computing the multipoles of the $NN$ numerator and $RR$ denominator,
one can easily transform to the multipoles of $\xi_{\rm aniso}$.
For the 3PCF, we have shown that the SE15 estimator for the Legendre
decomposition of the 3PCF can be computed with FTs.  

In all cases, the resulting algorithm scales only linearly with the 
number of survey objects (or random samples) and only $\mathcal{O}(N_g \ln
N_g)$ with the grid size.  For some important applications, this
is faster than the $\mathcal{O}(N n V_{R_{\rm max}})$ scaling of the pair-counting
methods (and the SE15 3PCF method).  The speed advantages are maximized when one considers
larger separations, higher number densities, and coarser grids.

However, Fourier methods do introduce artifacts due to grid resolution,
the level of which will depend on the ratio of the grid spacing to
the radial bin width being used.  For example, in 
the BAO analysis of Anderson et al. (2014), 2PCF separation bins of $8h^{-1}$~Mpc were used.
One would then want the FT grid to be comfortably smaller than this.
It is worth noting that the FT artifacts would be reduced if one used
radial separation bins with smoother edges instead of the traditional
tophats.  Smoother bins are acceptable for science applications such
as BAO, which is a smooth feature itself, and indeed are numerically
more stable for the spherical Bessel transforms needed to form model 2PCF
from theoretical power spectra.  One might proceed by tuning the bins 
to have negligible support beyond wavenumber $k\approx 0.4h$~Mpc$^{-1}$, 
where the acoustic oscillations have been damped away,
while choosing an FT grid scale of 3--4$h^{-1}$~Mpc so as to place
the Nyquist frequency comfortably above that smoothing scale.

We expect that an important application of these methods is to the
calculation of correlation functions from mock catalogs.  The
computation of covariance matrices now is commonly performed by
repeating the calculation on hundreds or thousands of mock catalogs.
This dominates the computational effort of the cosmology analysis.
But it is a place where making a mild sacrifice in the accuracy of
the pair count may be acceptable to gain the speed FTs offer. An interesting aspect of these FT approximations is that
they should converge to the pair-counting answer as the grid size
increases.  One might opt to compute mocks with a reasonably efficient
grid, accepting an error that is small compared to the survey
variations, while still processing the actual data with a finer grid
or an explicit pair-counting code.

While we were preparing this work for submission, Bianchi et al.
(2015) and Scoccimarro (2015) posted pre-prints suggesting use of FTs to compute the
anisotropies of the large-scale power spectrum about the line-of-sight,
using respectively the Yamamoto et al. (2006) estimator and a newly constructed estimator. Scoccimarro also uses FTs to estimate the redshift-space bispectrum. The core mathematical approach of these works is the same as what we present in Section \ref{sec:aniso_corr} for
the anisotropic 2PCF.  In detail, they present their results as
polynomial expansions of Legendre polynomials, whereas we expand
in spherical harmonics.  The spherical harmonics can be directly
translated to polynomials when computing (SE15 Section 2), and extensions to higher
multipoles are likely more convenient to track with $Y_{\ell m}$.
We note that the very large computational advantage reported in 
Bianchi et al. (2015) is specific to the power spectrum, which otherwise
required summing over all pairs of survey objects.  For the 2PCF,
common methods only need to count pairs within $\Vmax$,
so the FT advantage over pair counting is more modest for realistic grid sizes.

The coming generation of large galaxy surveys will stress our
computational resources not simply because of the survey size but also
because of the drive for increasing analysis accuracy, which manifests
itself in larger numbers of mock catalogs and analysis variations.
We believe that Fourier methods such as those presented here offer an important means of enhancing the computational speed of future cosmological analyses.

\section*{Acknowledgments}
The authors gratefully acknowledge useful discussions with Blakesley Burkhart, Charles-Antoine Collins Fekete, Doug
Finkbeiner, Nikhil Padmanabhan, Stephen Portillo, and Martin White. This material
is based upon work supported by the National Science Foundation
Graduate Research Fellowship under Grant No. DGE-1144152 and by the Department of Energy under grant DE-SC0013718.

\section*{References}

\noindent Alam S. et al., 2015, preprint (arXiv:1501.00963) %SDSS DR11 and DR12 paper. http://adsabs.harvard.edu/cgi-bin/bib_query?arXiv:1501.00963
%done.
\\
\noindent Alcock C. \& Paczy\'nski B., 1979, Nature 281, 358-359%http://www.nature.com/nature/journal/v281/n5730/abs/281358a0.html
%done
\\
\noindent Anderson L. et al., 2014, MNRAS, 441, 1, 24-62%http://adsabs.harvard.edu/cgi-bin/bib_query?arXiv:1312.4877
%done

\hangindent=1.5em
\hangafter=1
\noindent Arfken G.B., Weber H.J. \& Harris F.E., 2013, Mathematical Methods for Physicists: Academic Press, Waltham, MA
%done.

\hangindent=1.5em
\hangafter=1
\noindent Bernardeau F., Colombi S., Gazta\~{n}aga E., Scoccimarro R., 2002, Phys. Rep., 367, 1

\hangindent=1.5em
\hangafter=1
\noindent Bianchi D., Gil-Mar\'in H., Ruggeri R. \& Percival W.J., 2015, MNRAS 453, 1, L11-L15%http://adsabs.harvard.edu/cgi-bin/bib_query?arXiv:1505.05341

\hangindent=1.5em
\hangafter=1
\noindent Cabr\'e A. \& Gazta\~{n}aga E., 2009, MNRAS, 393, 1183-1208%http://mnras.oxfordjournals.org/content/393/4/1183.full.pdf
%uses multipoles, see eqn 16.

\hangindent=1.5em
\hangafter=1
\noindent Chen G. \& Szapudi I., 2005, ApJ, 635:743-749%http://arxiv.org/abs/astro-ph/0508316
%done.

\hangindent=1.5em
\hangafter=1
\noindent Chuang C.-H., Wang Y., 2012, MNRAS, 426, 226-236%http://adsabs.harvard.edu/cgi-bin/bib_query?arXiv:1102.2251
%method for getting H and D_A from 2-D 2PCF. they use multipoles; see eqns 6-8 (up to l=4).

\hangindent=1.5em
\hangafter=1
\noindent Chuang C.-H . \& Wang Y., 2013a, MNRAS 435, 1, 255-262%http://adsabs.harvard.edu/cgi-bin/bib_query?arXiv:1209.0210
%Modelling the anisotropic two-point galaxy correlation function on small scales and single-probe measurements of H(z), DA(z) and f(z)σ8(z) from the Sloan Digital Sky Survey DR7 luminous red galaxies
%involves RSD, used LAsDamas, applied to SDSS DR7 LRGs, monopole and quadrupole of correlation function, 25<r120 hinv Mpc.

\hangindent=1.5em
\hangafter=1
\noindent Chuang C.-H. \& Wang Y., 2013b, MNRAS, 431, 3, 2634-2644%http://adsabs.harvard.edu/cgi-bin/bib_query?arXiv:1205.5573
%title is very relevant, but lots of overlap with Chuang & Wang 2012.

\hangindent=1.5em
\hangafter=1
\noindent Chuang C.-H. et al., 2013, preprint (arXiv:1312.4889) %http://adsabs.harvard.edu/cgi-bin/bib_query?arXiv:1312.4889
%BOSS DR-11 CMASS, uses multipoles, l=0, l=2.

\hangindent=1.5em
\hangafter=1
\noindent Coil A., 2013, ``The Large-Scale Structure of the Universe'' in Planets, Stars and Stellar Systems Vol. 6, eds. Oswalt T.D. \& Keel W.C.: Springer, Dordrecht %http://adsabs.harvard.edu/cgi-bin/bib_query?arXiv:1202.6633
%done

\hangindent=1.5em
\hangafter=1
\noindent Cooley J.W. \& Tukey J.W., 1965, Math. Comput. 19: 297-301

\hangindent=1.5em
\hangafter=1
\noindent Feldman H.A., Kaiser N. \& Peacock J.A., 1994, ApJ, 426, 23

\hangindent=1.5em
\hangafter=1
\noindent Gazta\~{n}aga E., Cabr\'e A. \& Hui L., 2009, MNRAS, 399, 3, 1663-1680%http://adsabs.harvard.edu/cgi-bin/bib_query?arXiv:0807.3551
%cite for anisotropic BAO signature.
%done.

\noindent Guo H. et al., 2014, MNRAS, 449, 1, L95-L99%http://adsabs.harvard.edu/cgi-bin/bib_query?arXiv:1409.7389
%for 3PCF
%done

\hangindent=1.5em
\hangafter=1
\noindent Hamilton A.J.S, 1998, in The Evolving Universe: Selected Topics on Large-Scale Structure and on the Properties of Galaxies, Dordrecht: Kluwer%http://adsabs.harvard.edu/cgi-bin/bib_query?arXiv:astro-ph/9708102
%done

\hangindent=1.5em
\hangafter=1
\noindent Hockney R.W. \& Eastwood J.W., 1981, Computer simulations using particles: Mc Graw-Hill, New York%http://adsabs.harvard.edu/abs/1981csup.book.....H
%for TSC, cited in Jing 2005
%done

\noindent Jain B. et al., 2015, preprint (arXiv:1501.07897v2)%''The whole is greater than the sum of its parts''---paper I presented on at group meeting a month or so ago.
%done

\noindent Jing Y.P., 2005, ApJ 620, 2, 559-563 %http://arxiv.org/pdf/astro-ph/0409240v2.pdf 
%fixing effects of TSC on power spectrum 
%done

\noindent Kaiser N., 1987, MNRAS, 227, 1-21%http://adsabs.harvard.edu/abs/1987MNRAS.227....1K
%done

\noindent Kayo I. et al., 2004, PASJ, 56, 415
%done

\noindent Landy S.D. \& Szalay A.S., 1993, ApJ, 412, 1 %http://adsabs.harvard.edu/abs/1993ApJ...412...64L
%done.

\hangindent=1.5em
\hangafter=1
\noindent Laureijs R. et al., 2011, Euclid Definition Study Report, preprint (arXiv:1110.3193)
%done.

\noindent Levi M. et al., 2013, preprint (arXiv:1308.0847)%http://adsabs.harvard.edu/cgi-bin/bib_query?arXiv:1308.0847
%done

\hangindent=1.5em
\hangafter=1
\noindent LSST Dark Energy Science Collaboration, 2012, preprint (arXiv:1211.0310)%http://adsabs.harvard.edu/cgi-bin/bib_query?arXiv:1211.0310
%done

\hangindent=1.5em
\hangafter=1
\noindent McBride C., Connolly A. J., Gardner J. P., Scranton R., Newman J., Scoccimarro R., Zehavi I., Schneider D. P., 2011a, ApJ, 726, 13
%http://adsabs.harvard.edu/abs/2011ApJ...726...13M
%done

\hangindent=1.5em
\hangafter=1
\noindent McBride K., Connolly A. J., Gardner J. P., Scranton R., Scoccimarro R., Berlind A., Marin F., Schneider D. P., 2011b, ApJ, 739, 85
%http://adsabs.harvard.edu/abs/2011ApJ...739...85M
%done

\noindent Okumura T. \& Jing Y.P., 2011, ApJ 726, 1, 5%http://adsabs.harvard.edu/cgi-bin/bib_query?arXiv:1004.3548
%they did look at xi_2/xi_0 to get beta though this is buried in conclusion. discuss systematics of growth factor from RSD.

\hangindent=1.5em
\hangafter=1
\noindent Peebles P. J. E., 1980, The Large-Scale Structure of the Universe: Princeton University Press, Princeton

\hangindent=1.5em
\hangafter=1
\noindent Percival W.J. et al., 2014, MNRAS 439, 3, 2531-2541%http://adsabs.harvard.edu/cgi-bin/bib_query?arXiv:1312.4841

\hangindent=1.5em
\hangafter=1
\noindent Press W., Teukolsky S., Vetterling W.T. \& Flannery B.P., 2007, Numerical Recipes: The Art of Scientific Computing: Cambridge University Press, Cambridge 

\noindent Raccanelli A. et al., 2013, MNRAS, 436, 1, 89-100%http://adsabs.harvard.edu/abs/2013MNRAS.436...89R
%focuses on RSD, l=0, 2 of 2PCF, corrections due to geometry, wider range of scales. SDSS-II DR7.
%appendix 1 has wide-angle effect discussion, interesting stuff on tripolar spherical harmonics, Szapudi and Papai.

\noindent Reid B.A. et al., 2012, MNRAS, 426, 4, 2719-2737%http://adsabs.harvard.edu/cgi-bin/bib_query?arXiv:1203.6641 
%xi_0 and xi_2 from BOSS DR9.

\hangindent=1.5em
\hangafter=1
\noindent Ross A.J., Percival W.J. \& Manera M., 2015, preprint (arXiv:1501.05571)%http://adsabs.harvard.edu/cgi-bin/bib_query?arXiv:1501.05571
%information from corr fn and power spec contained in multipoles

\hangindent=1.5em
\hangafter=1
\noindent Scoccimarro R., 2015, preprint (arXiv:1506.02729)%http://adsabs.harvard.edu/abs/2015arXiv150602729S

\hangindent=1.5em
\hangafter=1
\noindent Samushia L., Percival W.J. \& Raccanelli A., 2012, MNRAS 420, 3, 2102-2119%http://adsabs.harvard.edu/cgi-bin/bib_query?arXiv:1102.1014
%Interpreting large-scale redshift-space distortion measurements
%studies sources of uncertainty in RSD measurements using LasDamas--found important at 20% level; lots of good references. should we include Scoccimarro 2004 (phenomenological corrections to Kaiser formula) and Jackson 1972 (FoG) in our inline cite?
%measures fsigma8 with SDSS DR7.
%mentioned in Chuang 2012 as showing wide angle effects are small.

\noindent S\'anchez A.G. et al., 2013, MNRAS, 440, 3, 2692-2713
%http://adsabs.harvard.edu/cgi-bin/bib_query?arXiv:1312.4854
%uses multipoles, DR10 and 11

%Matsubara T., 
%http://arxiv.org/pdf/0807.1733v4.pdf
%PT model for non-linear effects in 2 pt function--about biasing.

%http://arxiv.org/pdf/0808.0003v2.pdf Percival & White 2008 is useful though not relevant for this.

\hangindent=1.5em
\hangafter=1
\noindent Scoccimarro R., 2015, preprint (arXiv:1506.02729v1)%http://adsabs.harvard.edu/cgi-bin/bib_query?arXiv:1506.02729

\hangindent=1em
\hangafter=1
\noindent Slepian Z. \& Eisenstein D.J., 2015, preprint (arXiv:1506.02040)%http://adsabs.harvard.edu/cgi-bin/bib_query?arXiv:1506.02040

\noindent Spergel D. et al., 2013, preprint (arXiv:1305.5422)%http://adsabs.harvard.edu/cgi-bin/bib_query?arXiv:1305.5422
%done

\hangindent=1.5em
\hangafter=1
\noindent Szapudi I., 2005, in Data Analysis in Cosmology, eds. Martinez V.J., Martinez-Gonzalez E., Pons-Borderia M.J. \& Saar E., Springer-Verlag Lecture Notes in Physics%http://adsabs.harvard.edu/abs/2005astro.ph..5391S
%done

\hangindent=1.5em
\hangafter=1
\noindent White M., Reid B.A., Chuang C.-H., Tinker J.L., McBride C.K., Prada F. \& Samushia L., 2015, MNRAS, 447, 1, 234-245%http://adsabs.harvard.edu/cgi-bin/bib_query?arXiv:1408.5435
%uses multipoles; eqn. 2 and 3, l=0,2.

\hangindent=1.5em
\hangafter=1
\noindent Xu X., Cuesta A.J., Padmanabhan N., Eisenstein D.J. \& McBride C.K., 2013, MNRAS, 431, 3, 2834-2860
%http://arxiv.org/pdf/1206.6732v2.pdf
%useful

\hangindent=1.5em
\hangafter=1
\noindent Yamamoto K., Nakamichi M., Kamino A., Bassett B. A., Nishioka
H., 2006, PASJ, 58, 93
%Done

\noindent Zheng Z., 2004, ApJ 614, 527%paper of person who emailed us; I looked at it quickly and it looks fine.
%done

\end{document}